\documentclass[useAMS,usenatbib,referee]{biom}

\usepackage{amssymb, amsmath}
\usepackage{appendix}
\usepackage{graphicx}
\usepackage{bm}

\usepackage[ruled,vlined]{algorithm2e}
\usepackage{booktabs}
\usepackage{hyperref}
\usepackage{threeparttable}
\usepackage[figuresright]{rotating}

\title[Bayesian modeling of co-occurrence microbial interaction networks]{Bayesian modeling of co-occurrence microbial interaction networks}

\author{Tejasv Bedi$^{1}$, Bencong Zhu$^{1,2}$, Michael L. Neugent$^{3}$, Kevin C. Lutz$^{4}$, \\
\textbf{Nicole J. De Nisco$^{\bm{3}}$, and Qiwei Li}$^{\bm{1,*}}$\email{qiwei.li@utdallas.edu} \\
$^{1}$Department of Mathematical Sciences, The University of Texas at Dallas, Richardson, Texas, U.S.A \\
$^{2}$Department of Statistics, The Chinese University of Hong Kong, Hong Kong\\
$^{3}$Department of Biological Sciences, The University of Texas at Dallas, Richardson, Texas, U.S.A\\
$^{4}$O'Donnell School of Public Health, The University of Texas Southwestern Medical Center, Dallas, Texas, U.S.A
}

\begin{document}


\date{{\it Received October} 2007. {\it Revised February} 2008.  {\it
Accepted March} 2008.}



\pagerange{\pageref{firstpage}--\pageref{lastpage}} 
\volume{64}
\pubyear{2023}
\artmonth{May}


\doi{10.1111/j.1541-0420.2005.00454.x}


\label{firstpage}


\begin{abstract}
The human body consists of microbiomes associated with the development and prevention of several diseases. These microbial organisms form several complex interactions that are informative to the scientific community for explaining disease progression and prevention. Contrary to the traditional view of the microbiome as a singular, assortative network, we introduce a novel statistical approach using a weighted stochastic infinite block model to analyze the complex community structures within microbial co-occurrence microbial interaction networks. Our model defines connections between microbial taxa using a novel semi-parametric rank-based correlation method on their transformed relative abundances within a fully connected network framework. Employing a Bayesian nonparametric approach, the proposed model effectively cluster taxa into distinct communities while estimating the number of communities. The posterior summary of the taxa’s community membership is obtained based on the posterior probability matrix, which could naturally solve the label switching problem. Through simulation studies and real-world application to microbiome data from postmenopausal patients with recurrent urinary tract infections, we demonstrate that our method has superior clustering accuracy over alternative approaches. This advancement provides a more nuanced understanding of microbiome organization, with significant implications for disease research.
\end{abstract}

%



\maketitle


%

\section{Introduction}
\label{s:intro}

Microbial ecological interactions are fundamental in shaping microbiome functionality and host health by creating complex communities where microbes coexist \citep{antwis2017fifty}. Extensive research highlights the importance of these interactions in maintaining the equilibrium of a healthy microbiome \citep{huttenhower486giglio, faust2012microbial}. Furthermore, a significant body of literature emphasizes the necessity of elucidating microbial interactions to comprehend the microbiome's impact on disease progression and vulnerability \citep[see e.g.,][]{frank2011investigating,marchesi2011towards,karlsson2013gut,shreiner2015gut,halfvarson2017dynamics,hall2019co}. Hence, there is an escalating demand for the development of network-based methods to investigate patterns of co-occurrence or co-exclusion among microbial taxa \citep{hall2019co}, which are essential for advancing our understanding of microbiome dynamics.

A microbiome co-occurrence network is represented as a graph, comprising nodes and edges, where each node corresponds to a taxon and each edge signifies a significant association between two taxa. An inferred microbiome co-occurrence network facilitates the characterization of taxon-taxon associations, uncovering their hidden attributes, mechanisms, and structures \citep{jiang2020harmonies}. For a comprehensive review of the statistical methods employed in the analysis of microbiome networks, including co-occurrence networks, we recommend the survey by \cite{lutz2022survey}.

Community detection is one of the fundamental problems in network analysis, with the stochastic block model (SBM) emerging as a widely used unsupervised learning technique. The foundational mathematics of the SBM were first laid out by \cite{holland1983stochastic}, and then later, \cite{guimera2009missing} notably applied the SBM for model selection in estimating network structures. SBM partitions the nodes of a diverse network into $K$ distinct, homogeneous communities or blocks, based on their connectivity patterns \citep{mcdaid2013improved, aicher2015learning, lee2019review}. \cite{lutz2023generalized} provide a non-exhaustive catalog with descriptions of a variety of existing methods for users in \texttt{R}, \texttt{Python}, and \texttt{C++}. SBM has found recent application in a variety of biological networks including taxa-taxa communities in the human gut microbiome associated with disease development \citep{hall2019co} and the human urinary microbiome related to recurrent urinary tract infections (rUTI) \citep{lutz2023generalized}, as well as protein-protein interactions related to pancreatic cancer \cite{stanley2019stochastic}, among others.

Initially, SBM was limited to binary or unweighted networks, capturing merely the presence or absence of edges without considering their weights. However, networks with edges weighted by continuous values offer a richer depiction of the network structure \citep{aicher2015learning}. The development of the weighted SBM (WSBM) by \cite{aicher2013adapting} marked a significant advancement, leading to numerous studies \citep[see e.g.,][]{faskowitz2018weighted,peixoto2018nonparametric,ahn2018hypergraph, xu2020optimal, chen2022community} that have broadened the scope and refinement of SBM applications. Despite their proven capability in community detection, these WSBM methodologies encounter notable challenges, particularly when applied to microbiome co-occurrence networks. A common limitation is the prerequisite for a predetermined number of communities $K$ or reliance on certain criteria to deduce the optimal $K$ during post-analysis \citep[see e.g.,][]{biernacki2000assessing,aicher2015learning,lutz2023generalized}, with some strategies incorporating the reversible jump Monte Carlo Markov chain algorithm for simultaneous estimation of $K$ and other parameters \citep[see e.g.,][]{wyse2012block, ludkin2020inference}. Moreover, existing methods are not directly applicable to microbiome data analysis due to their failure to address key characteristics of taxonomic abundance data such as high-dimensionality, zero-inflation, over-dispersion, and compositionality \citep{cullen2020emerging}. This oversight can result in loss of information and increased bias in inference. Consequently, tailored WSBM models accounting for those challenging characteristics are required to perform community detection on a microbiome co-occurrence network. 

In this paper, we introduce a Bayesian weighted stochastic infinite block model (WSIBM) that is designed to detect communities within a weighted network, estimate the parameters for each community, and infer the number of communities $K$ using a nonparametric Bayesian method. To fit the model, we construct a blocked Gibbs sampler \textit{via} stick-breaking priors \citep{ishwaran2001gibbs}, which facilitates the estimation of $K$ through a truncated Dirichlet processes (DP). Our findings reveal that the WSIBM surpasses a standard Bayesian WSBM when the latter is constrained by a predefined $K$. To address specific challenges presented by microbiome data, we implement a modified centered log-ratio (MCLR) transformation \citep{yoon2019microbial} to correct for zero-inflation and non-linearity, and apply a recently developed semi-parametric rank-based (SPR) correlation method \citep{yoon2020sparse} to construct the weighted microbiome co-occurrence network. Furthermore, we explore the community structure of the urinary microbiome in postmenopausal women suffering from rUTI. This investigation marks the first study of the urinary microbiome's weighted co-occurrence network structure. The results of our analysis provide the foundation for future studies related to the urinary microbiome and human health.

\begin{figure}
    \centering
    \includegraphics[width = 1\linewidth]{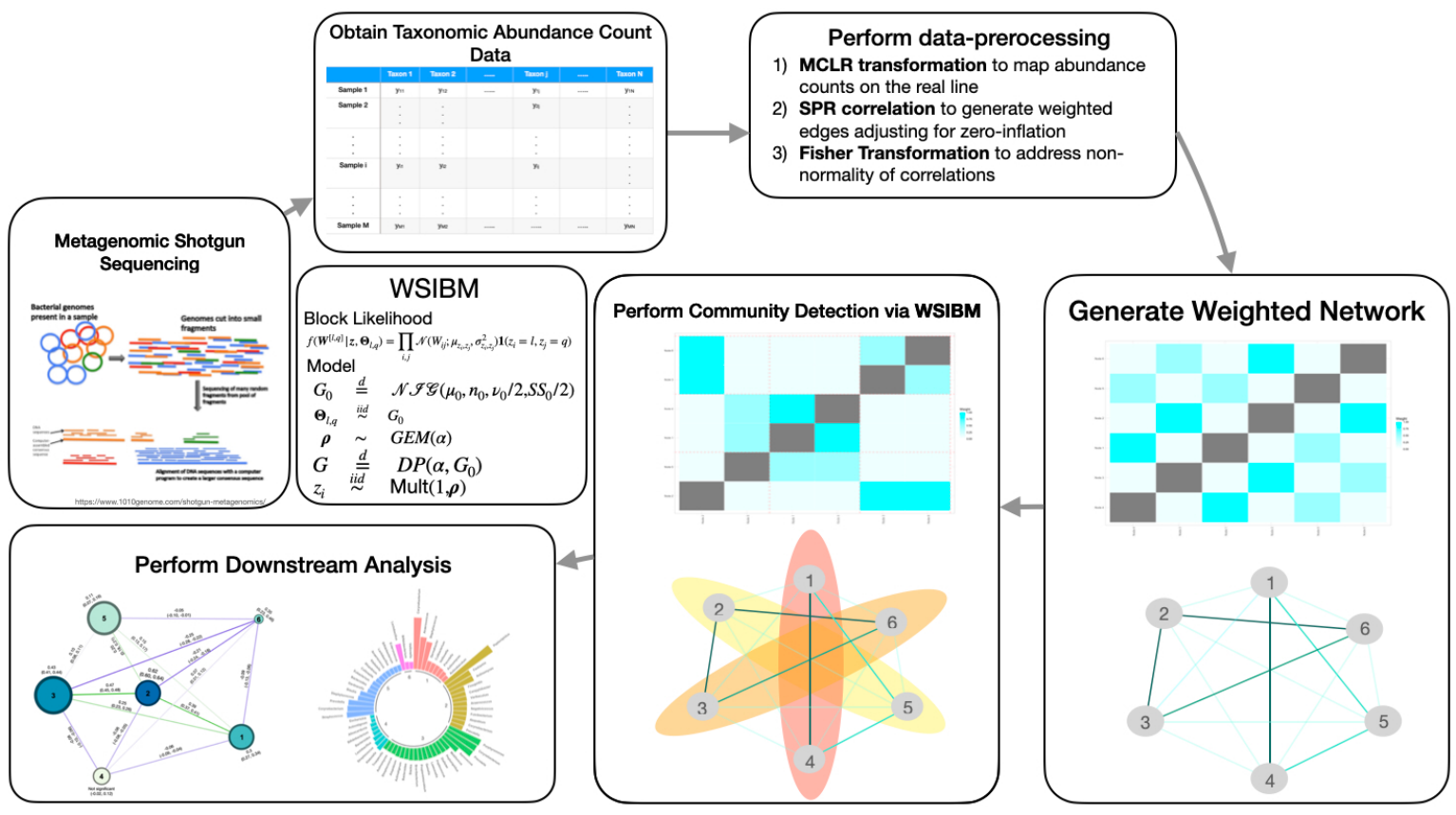}
    \caption{A flowchart to summarize the entire modeling pipeline}
    \label{fig:workflow}
\end{figure}

The workflow of our model is presented in Figure \ref{fig:workflow}. The paper proceeds as follows. Section \ref{s:background} provides an overview of the necessary background information and the data preprocessing steps required prior to applying our model. In Section \ref{s:model}, we elaborate on the development of both WSBM and WSIBM, including the model fitting process using the blocked Gibbs sampler and the truncated DP. Section \ref{s:inf} delves into the estimation procedure where we derive both point and interval estimates for several key parameters. We then proceed to validate our model through simulation studies presented in Section \ref{s:sim}. Community detection using a real microbiome dataset from postmenopausal women afflicted with rUTI is discussed in Section \ref{s:res}. We conclude the paper in Section \ref{s:conclusion} with a discussion on the findings and potential avenues for future research.

\section{Background Material}
\label{s:background}
\subsection{Weighted Graph}
A directed weighted random graph can be denoted as $\mathcal{G}=(\mathcal{V},\mathcal{E}, \mathcal{W})$ where $\mathcal{V}$ is a set of nodes with $|\mathcal{V}| = n$, $\mathcal{E} = \{0, 1\}^{n \times n}$ is a set indicating the presence or absence of an edge between the nodes, and $\mathcal{W} = \mathbb{R}^{n \times n}$ is the weighted adjacency matrix. We further make some assumptions that reduce model complexity to allow direct application to microbiome data. Let $\bm{E} = \{E_{jj'}\}_{n \times n}$ and $\bm{W} = \{W_{jj'}\}_{n \times n}$ be random binary and weighted adjacency matrices, respectively, $\forall(j,j')\in \{1,\ldots,n\}$. Then, the assumptions are as follows: (1) the graph is fully connected (i.e., $E_{jj'} = 1$,  $\forall(j,j')\in \mathcal{V}$, $j \neq j'$), (2) the graph is undirected (i.e., $E_{jj'} = E_{j'j}$, $W_{jj'} = W_{j'j}$, $\forall(j,j')\in \mathcal{V}$, $j \neq j'$) and, (3) the graph has no self-loops (i.e., $E_{jj} = W_{jj} = 0$, $\forall j \in \mathcal{V}$). Now, having a fully connected graph from (1), we only observe variability in edge weights $W_{jj'}$ allowing us to ignore the edge presence adjacency matrix $E_{jj'}$. Secondly, from (2) and (3), $\bm{W}$ is symmetric with diagonal entries set to zero. For instance, let each node represent the $j$-th taxon in a microbiome network that has $n$ taxa. Then, $W_{jj'}\in \boldsymbol{W}$ denotes the value or weight assigned to the edge between taxa $j$ and $j'$ where $\boldsymbol{W}$ is the weighted adjacency matrix. Here, $W_{jj'}$ is a continuous value that measures the association between a pair of taxa. 

Next, we discuss the data preprocessing steps that map the taxonomic abundance data to the weighted adjacency matrix.

\subsection{Weighted Graph Construction for Taxonomic Abundance Data}
\label{data_pre}
 Let $\bm{Y} \in [0, 1]^{m \times n}$ be a matrix of relative 
taxonomic abundance data with $m$ samples and $n$ taxa such that 
$\bm{y}_i = (y_{i1},\hdots, y_{in})$ is a vector of relative 
abundances for the $i$-th sample where $i=1,\ldots,m$. Each of the $m$ samples have compositionality and, hence, belong to a $n$-dimensional simplex space given by
\begin{equation*}
    \mathcal{S}^{n - 1} \quad = \quad \left\{ (y_{i1},\hdots, y_{in}) : y_{ij} \geq 0, \text{ and } \sum_{j = 1}^n y_{ij} = 1 \right\}. \label{eq1}
\end{equation*}
There are $\binom{n}{2}$ number of possible interactions between the 
$n$ taxa. We may obtain the correlation matrix $\bm{R} = 
[R_{jj'}]_{n\times n}$ by estimating the pairwise correlation coefficients of 
$\binom{n}{2}$ pairs of taxa. However, we first remove 
compositionality by applying the central log ratio (CLR) transformation 
\cite{aitchison1984statistical,aitchison1982statistical} 
$\mathcal{C}: \mathcal{S}^{n - 1} \longrightarrow U\subset \mathbb{R}^{n - 1}$ on $\bm{y}_i$ such that 
\begin{equation*}
\mathcal{C}(\bm{y}_i) \quad=\quad \log\left[\frac{y_{i1}}{g(\bm{y}_i)},\hdots, 
\frac{y_{in}}{g(\bm{y}_i)} \right] \label{eq2}
\end{equation*}
where $g(\bm{y}_i) = (\prod_{j = 1}^n y_{ij})^{1/n}$ is the geometric mean of the $i$-{th} sample. To 
avoid dividing by zeros, we must add a pseudo value to the data and re-normalize it to maintain compositionality before applying the CLR 
transformation. However, the choice of pseudo value is arbitrary and masks the zeros, which means that zeros and non-zeros in the data may be treated equally. In order to avoid adding a pseudo value and to preserve the zeros in the data, \cite{yoon2019microbial} proposed a modified CLR (MCLR) transformation given by $\Tilde{\mathcal{C}}(\cdot)$ given by
\begin{equation*}
    \Tilde{\mathcal{C}}(\bm{y}_i) \quad=\quad \left[0, \hdots, 0, \log\left(\frac{y_{i(j^{*} + 1)}}{\Tilde{g}(\bm{y}_i)}\right) + \epsilon, \hdots, \log\left(\frac{y_{in}}{\Tilde{g}(\bm{y}_i)}\right) + \epsilon \right] \label{eq3}
\end{equation*}
where $\Tilde{g}(\bm{y}_i) = (\prod_{j = j^{*} + 1}^n y_{ij})^{1/(n - j^{*})}$ is the geometric mean of only the non-zero relative abundances with the adjustment factor $\epsilon = |\min_{ij}\log(y_{ij}/\Tilde{g}(\bm{y}_i))| + 1$ to ensure all non-zero values are strictly positive.

Once we apply the MCLR transformation to ensure that the relative abundances are free from the compositionality restriction, we evaluate taxa-taxa associations \textit{via} measures of correlation to compute the correlation matrix $\bm{R}$. One may compute pairwise Pearson or rank based Spearman correlation as a measure of $\bm{R}$. However, these methods cannot account for sparsity and excessive zeros. Correlation and partial correlation structure are inferred by various statistical methods tailored for microbiome network analysis. Several statistical methods for microbiome-related network analysis include SparCC (Sparse Correlations for Compositional data) \cite{friedman2012inferring}, CCLasso (Correlation inference for Compositional data through Lasso) \cite{fang2015cclasso}, and REBACCA (Regularized Estimation of the BAsis Covariance based on Compositional dAta) \cite{ban2015investigating}, SpiecEasi (SParse InversE Covariance Estimation for Ecological Association Inference) \cite{kurtz2015sparse}, HARMONIES (Hybrid Approach foR MicrobiOme Network Inferences \textit{via} Exploiting Sparsity) \cite{jiang2020harmonies} and SPRING (Semi-Parametric Rank-based approach for INference in Graphical model)  \cite{yoon2019microbial}.

SPRING uses novel SPR estimators of correlation and partial correlation for relative abundance data with the MCLR transformation. The semi-parametric model in SPRING combines a truncated Gaussian copula graphical model with rank-based partial correlation to construct a sparse network. The latent Gaussian copula model proposed by \cite{fan2017high} naturally addresses most of our concerns as it directly applies to undirected graphs and estimates the covariance matrix $\bm{\Sigma}$ \textit{via} a multivariate Gaussian distribution while relaxing the normality assumption. Microbiome data requires this flexibility as it is non-linear and may be heavily skewed. However, the model may not have the framework to account for excessive zeros. \cite{yoon2019microbial, yoon2020sparse} address this concern by deriving new bridge functions that connect $\bm{\Sigma}$ to the Kendall rank correlation coefficient (i.e., Kendall's $\tau$) and compute a rank-based estimator $\hat{\bm{R}}$ while accounting for excessive zeros in the data. The \texttt{SPRING} package \citep{yoon2019microbial} in \texttt{R} was used to estimate the correlation matrix $\bm{R}$.

Finally, we apply the Fisher transformation $\mathcal{F}:[-1, 1] \longrightarrow \mathbb{R}$ on each 
$R_{jj'}$ to obtain a weight matrix $\bm{W} = [W_{jj'}]_{n\times n}$ where
\begin{equation*}
    W_{jj'} \quad=\quad \mathcal{F}(R_{jj'}) \quad=\quad \frac{1}{2} \ln \left(\frac{1 + 
    R_{jj'}}{1 - R_{jj'}}\right) \label{eq4}.
\end{equation*}
The final transformation step maps the correlations to the real number line for the sake of consistency in the distributional assumptions of the Weighted Stochastic Block Model (WSBM) and Weighted Stochastic Infinite Block Model (WSIBM) proposed in the next section. The Fisher transformation stretches out the extreme correlations even further without affecting small to moderate correlations and results in a bell-shaped distribution for the correlation density in the weight matrix domain $\bm{W}$.

\section{Model}
In this section, we first explore the weighted stochastic block model (WSBM) as a finite block model having a fixed number of communities $K$. We further extend our model to a weighted stochastic infinite block model (WSIBM) that automatically infers $K$ \textit{via} a stick-breaking process. 
\label{s:model}
\subsection{Finite Block Model}
The weighted stochastic block model (WSBM) generates a pairwise 
interaction network of $n$ nodes defined by $\bm{W}$ where $W_{jj'}\in\bm{W}$ 
defines a weighted edge between the $j$-th and $j'$-th nodes.     Each node belongs to one of the $K$ latent groups or 
communities. Further, we assume a complete and undirected network 
having $\binom{n}{2}$ edges in total with no self loops $(j \neq j')$. 
The community membership of each node is defined by a latent community
allocation vector $\bm{z}$ where $ z_{j} \in \mathbb{N}^+$ and $n_k = \sum_{j = 1}^n I(z_j = 
k)$ is the total number of vertices belonging to group $k$ where $k=1,\ldots,K$ and $I(\cdot)$ is the  indicator function. Having 
fixed $K$ communities in total, we define a block as a sub-matrix 
$\bm{W}^{[l, q]} \subset \bm{W}$ where $\bm{W}^{[l, q]} = \{W_{jj'}: 
z_j = l, z_{j'} = q\}$ for all pairs $(l,q) \in \{1,\hdots,K\}$. Here, 
the diagonal blocks $\bm{W}^{[l, l]}$ for $ l \in \{1,\hdots,K\}$ 
have within-community interactions. While, the off-diagonal blocks $\bm{W}^{[l, q]}$ for $ l \neq q$ have between-community interactions. We further assume $\bm{W}^{[l, q]} = \bm{W}^{[q, l]}$,  resulting in $K(K+1)/2$ blocks in total.

Each block $\bm{W}^{[l, q]}$ is characterized by block-specific mean and variance parameters denoted as $\bm{\Theta}_{l,q} = \{\mu_{l,q}, \sigma^2_{l,q}\}$, which define 
edge-weight mean and variance respectively. The data likelihood of each block conditional on the community 
allocation vector $\bm{z}$ and block parameters $\bm{\Theta}_{l,q}$ is
\begin{equation*}
\begin{aligned}
    f(\bm{W}^{[l,q]}|\bm{z}, \bm{\Theta}_{l,q}) &= \prod_{j,j'} {N}({W}_{jj'}; \mu_{z_{j},z_{j'}}, \sigma^2_{z_{j},z_{j'}}) I(z_j = l, z_{j'} = q) \\
    &= \prod_{(j,j') \in \mathcal{S}_{l,q}} {N}({W}_{jj'}; \mu_{l,q}, \sigma^2_{l,q}) \label{eq5}
\end{aligned}
\end{equation*}
where $\mathcal{S}_{l,q} = \{(j,j'): j \neq j', z_j = l, z_{j'} = q\}$ and
 $\bm{W}^{[l,q]} \subset \bm{W}$ $\forall(l,q) = 1,\hdots,K$.
To complete the model, we specify independent conjugate priors for block-specific parameters $\bm{\Theta}_{l,q} = \{\mu_{l,q}, \sigma^2_{l,q}\} \sim 
\text{NIG}(\mu_0, n_0, \nu_0/2, SS_0/2)$, which is a 
normal-inverse-gamma (NIG) distribution with fixed hyperparameters $\mu_0 
\in \mathbb{R}, n_0 > 0, \nu_0 > 0,$ and $SS_0 > 0$ prespecified for 
non-informative settings. Equivalently, we may construct the priors as
follows: $\mu_{l,q}|\sigma^2_{l,q} \sim {N}(\mu_0, 
\sigma^2_{l,q}/n_0)$ \text{ and } $\sigma^2_{l,q} \sim \text{IG}(\nu_0/2, SS_0/2)$.
Then, we can derive the posterior densities of block means and variances $\bm{\Theta}_{l,q}$ for all $(l,q) = 1,\hdots, K$. Due to conjugacy, we obtain the posterior densities of $\bm{\Theta}_{l,q}$ given $\bm{z}$ by
\begin{equation*}
    \begin{aligned}
        \pi(\mu_{l,q}|\bm{z}, \sigma^2_{l,q}, \bm{W}^{[l,q]}) \quad &\propto \quad {N}(\mu^{(p)}_{l,q},\; \sigma^2_{l,q}/n^{(p)}_{l,q})\\
        \pi(\sigma^2_{l,q}|\bm{z}, \bm{W}^{[l,q]}) \quad &\propto \quad  \text{IG}(\nu^{(p)}_{l,q}/2,\; SS^{(p)}_{l,q}/2)
    \end{aligned} \label{eq6}
\end{equation*}
where total number of edges $N_{l, q}$ and sample mean
$\overline{W}_{l, q}$ for block $\bm{W}^{[l,q]}$ are defined by
\begin{equation*}
        N_{l, q} = \begin{cases} 
       \binom{n_l}{2}, \quad \text{if  } l = q \\
       n_l n_q, \quad \text{otherwise}
       \end{cases} \quad \text{and} \quad \overline{W}_{l, q} = \frac{\sum_{(j,j')\in\mathcal{S}_{l, q}} W_{jj'}}{N_{l, q}},
\end{equation*}
respectively. Subsequently, the posterior parameters are $n_{l, q}^{(p)} = N_{l, q} + n_0$, $\nu_{l, q}^{(p)} = N_{l, q} + \nu_0$, $\mu_{l, q}^{(p)} = \frac{N_{l, q}\overline{W}_{l, q} + n_0\mu_0}{N_{l, q} + n_0}$ and $SS_{l, q}^{(p)} = SS_0 + \sum_{(j,j')\in\mathcal{S}_{l, q}} (W_{jj'} - \overline{W}_{l, q})^2 + \frac{n_0N_{l, q}}{N_{l, q} + n_0}(\overline{W}_{l, q} - \mu_0)^2$. 

Finally, we derive the posterior density of the latent community membership vector $\bm{z}$. Assigning a multinomial (Mult) prior on $\bm{z}$, which is written as $\bm{z}|\bm{\nu} \sim \text{Mult}(1; \tau_1,\hdots, \tau_K)$ such that each node $j = 1,\hdots, n$ takes a community label from $l = 1,\hdots,K$ with probability $\tau_l$. Furthermore, we assume a Dirichlet (Dir) prior on the vector of probabilities $\bm{\tau}|\bm{\eta} \sim \text{Dir}(\eta_1, \hdots, \eta_K)$ with fixed hyper-parameter $\bm{\eta}$. The setting $\bm{\eta} = (1, 1, \hdots, 1)^{\top}$ imposes a non-informative prior on $\bm{\tau}$, which is equivalent to assuming a discrete uniform prior on $\bm\tau$ with each cluster label $l = 1,\hdots,K$ having equal probability of selection. However, setting $\tau_l > 1$ for all $l$ encourages communities to have a similar number of nodes. On the other hand, setting $\tau_l < 1$ for all $l$ discourages communities from having a similar number of nodes. The posterior density of $z_j$ for all $j = 1, \hdots, n$ is 
\begin{equation*}
    \begin{aligned}
        \pi(z_j = l|\bm{\Theta},\bm{W}, \bm{\tau}) \quad &\propto \quad \left[\prod_{q = 1}^K f(\bm{W}^{[l, q]}|z_j = l, z_{j'} = q, \bm{\Theta}_{l, q})\right] \pi(z_j = l|\bm{\tau})\\
        \quad &\propto \quad \tau_l\left[\prod_{q = 1}^K f(\bm{W}^{[l, q]}|z_j = l, z_{j'} = q, \bm{\Theta}_{l, q})\right].
    \end{aligned} \label{eq7}
\end{equation*}
We then normalize the posterior densities of $z_j$ for each $j$ to obtain the exact probabilities of community membership for block $l$ by
\begin{equation*}
    p_{jl} = p(z_j = l|\bm{\Theta},\bm{W}, \bm{\tau}) = \frac{\pi(z_j = l|\bm{\Theta},\bm{W}, \bm{\tau})}{\sum_{q = 1}^{K}\pi(z_j = q|\bm{\Theta},\bm{W}, \bm{\tau})}. \label{eq8}
\end{equation*}
Finally, we sample the community label $z_j$ from a multinomial 
distribution given by 
\begin{equation*}
    z_j|\bm{\Theta}, \bm{W}, \bm{\tau} \quad \sim \quad \text{Mult}(1, p_{j1},\hdots, p_{jK}). \label{eq9}
\end{equation*}
Subsequently, we also update $\bm{\tau}$ \textit{via} Dirichlet-multinomial conjugacy given by
\begin{equation*}
    \begin{aligned}
        \pi(\bm{\tau}|\bm{z}, \bm{\eta}) \quad &\propto \quad \pi(\bm{z}|\bm{\tau}, \bm{\eta})\pi(\bm{\tau}|\bm{\eta})
        \quad \propto \quad \prod_{l = 1}^K \tau_l^{n_l + \eta_l}\\
        \bm{\tau}|\bm{z}, \bm{\eta} \quad &\sim \quad \text{Dir}(n_1 + \eta_1, \hdots, n_K + \eta_K)
    \end{aligned}.
\end{equation*}
We estimate $\bm{z}$, $\bm{\tau}$, and $\bm{\Theta}$ for a fixed number of communities $K$ \textit{via} a blocked Gibbs sampler that updates $\bm{z}$, $\bm{\tau}$, and $\bm{\Theta}$ successively in blocks. The model fitting procedure described above is summarized in Algorithm \ref{alg1}.

\SetKwComment{Comment}{/* }{ */}
\begin{algorithm}[ht!]
\caption{Blocked Gibbs Sampler to fit Finite Block Model} \label{alg1}
\KwData{$\bm{W}$; fix $\mu_0, SS_0, \nu_0, n_0, K, \bm{\eta}, iter$}
\textbf{Initialize: }$\bm{\Theta}, \bm{z}, \bm{\tau}$\\
\For{$t$ in $1:iter$}{
   \For{$j$ in $1:n$}{
      \For{$l$ in $1:K$}{
         $p_{jl}^{(t)} = \frac{\pi(z_j^{(t - 1)} = l|\tau_{l}^{(t - 1)}, \bm{\Theta}^{(t-1)}, \bm{W})}{\sum_{q = 1}^{K} \pi(z_{j}^{(t - 1)}  = q|\tau_{q}^{(t - 1)}, \bm{\Theta}^{(t-1)}, \bm{W})}$ \Comment*[r]{Update $\bm{z}$}
      }
      $z_{j}^{(t)}|\bm{\tau}^{(t - 1)}, \bm{\Theta}^{(t - 1)}, \bm{W} \sim \text{Mult}(1, p_{j1}^{(t)},\hdots,p_{jK}^{(t)})$ \;
   }
   $\bm{\tau}^{(t)}|\bm{z}^{(t)}, \bm{\eta} \sim  \text{Dir}(n_1^{(t)} + \eta_1, \hdots, n_K^{(t)} + \eta_K)$ \Comment*[r]{Update $\bm{\tau}$}
   \For{$(l, q)$ in $1:K$}{
      $(\sigma_{l,q}^{2})^{(t)}|\bm{z}^{(t)}, \bm{W}^{[l,q]} \sim  \text{IG}((\nu^{(p)}_{l,q})^{(t)}/2,\; (SS^{(p)}_{l,q})^{(t)}/2)$ \Comment*[r]{Update $\bm{\Theta}$}
      $\mu_{l,q}^{(t)}|\bm{z}^{(t)}, (\sigma^2_{l,q})^{(t)}, \bm{W}^{[l,q]} \sim {N}(\mu^{(p)(t)}_{l,q},\; (\sigma^2_{l,q})^{(t)}/n^{(p)(t)}_{l,q})$\;
   }
   \textbf{Store: } $\bm{\Theta}^{(t)}, \bm{z}^{(t)}$\;
}
\end{algorithm}

\subsection{Infinite Block Model}

In this section, we further generalize a weighted stochastic block model (WSBM) with a finite number of communities (i.e., $K$ is prespecified), to a model that assumes an infinite number of communities $(i.e., K \longrightarrow \infty)$ and is referred to as a weighted stochastic infinite block model (WSIBM). In practice, however, only finitely many communities 
that are realized by the data are estimated. The distribution of $K$ 
can be indirectly estimated by counting the number of non-empty 
community labels across all observations. In other words, $K$ is truncated to a finite number and all remaining empty communities are negligible. To perform this 
nonparametric estimation of $K$, our model employs the truncated stick-breaking 
construction of the Dirichlet Process (DP) \cite{ishwaran2001gibbs}. Let $G$ be an infinite dimensional discrete distribution over a
continuous space $\Omega$, $G_0 = \text{NIG}(\mu_0, n_0, \nu_0/2, 
SS_0/2)$ be a base measure, and $\alpha$ be a concentration parameter. Then, $G \overset{d}{=} \text{DP}(\alpha, G_0)$. Following \cite{ishwaran2001gibbs} with the truncated DP model, the WSIBM model is written as
\begin{equation*}
    \bm{W}|\bm{z}, \Theta \sim f(\bm{W}|\bm{z}, \Theta), \quad \bm{z}|\bm{\rho} \sim \sum_{k=1}^{K_{\max}} \rho_{k}\delta_{k}(\cdot), \quad \bm{\rho} \sim \text{GEM}(\alpha) \label{eq11}
\end{equation*}
where $\bm{\rho}$ contains the weights of the communities constructed \textit{via} the stick-breaking process. The distribution of $\bm\rho$ is called the Griffiths-Engen-McCloskey (GEM) distribution \cite{ewens1990population}, which is a special case of the DP. The idea behind this construction is to assume a stick of unit length, which will be successively broken into smaller pieces. The length of each piece represents the probability of the $k$-th community, which is denoted as $\rho_k$, and is defined by
\begin{equation}
    \rho_k = V_k \prod_{s < k} (1 - V_s) \text{ for }  k = 2,\hdots, K_{\max} - 1 \label{eq12}
\end{equation}
where $V_k \sim \text{Beta}(1, \alpha)$ and Equation \eqref{eq12} together form the GEM distribution. Truncation is imposed by assuming $V_{K_{\max}} = 1$ and 
$\rho_{K_{\max}} = 1 - \sum_{s = 1}^{K_{\max} - 1} \rho_s$. This process 
only allows a finite number of communities and has exponentially decreasing class weights as the number of communities increases. 

The posterior sampling can be summarized in two successive steps. To update each $\rho_k\in\bm\rho$, we first sample $V_k^*$ from the conditional posterior density   
\begin{equation*}
    V_{k}^{*}|\bm{z} \quad \sim \quad \text{Beta}\left(1 + n_k,\; \alpha + \sum_{l = k + 1}^{K_{\max}} n_l\right) \label{eq13}
\end{equation*}
where $n_{k} = \sum_{j=1}^{n}I(z_{j} = k)$. After obtaining the estimated probability $\bm{\rho}$, we then estimate $\bm{z}$, $\bm{\Theta}$, and $K$ \textit{via} a blocked Gibbs sampler. The model fitting procedure described above is summarized in Algorithm \ref{alg2}. 



\begin{algorithm}[ht!]
\caption{Blocked Gibbs Sampler to fit Infinite Block Model}\label{alg2}
\KwData{$\bm{W}$; fix $\mu_0, SS_0, \nu_0, n_0, K_{\max}, \alpha, iter$}
\textbf{Initialize: }$\bm{\Theta}, \bm{z}, \bm{\rho}$\\
\For{$t$ in $1:iter$}{
   \For{$k$ in $1:(K_{\max} - 1)$}{
    $V_{k}^{*(t)}|\bm{z}^{(t-1)} \sim \text{Beta}(1 + n_{k}^{(t-1)}, \alpha + \sum_{l = k + 1}^{K_{\max}} n_{j}^{(t-1)})$ \;
    $\rho_{k}^{(t)} = V_{k}^{*(t)} \prod_{s < k} (1 - V_{s}^{*(t)})$ \Comment*[r]{Update $\bm{\rho}$}
   }
   $V_{K_{\max}}^{(t)} = 1$, $\rho^{(t)}_{K_{\max}} = 1 - \sum_{k = 1}^{K_{max} - 1} \rho_{k}^{(t)}$\;
   \For{$j$ in $1:n$}{
      \For{$k$ in $1:K_{\max}$}{
         $p_{jk}^{(t)} = \frac{\rho_{k}^{(t)} \pi(z_{j}^{(t - 1)} = k|\bm{\Theta}^{(t-1)}, \bm{W})}{\sum_{l = 1}^{K_{\max}} \rho_{l}^{(t)} \pi(z_{j}^{(t - 1)}  = l|\bm{\Theta}^{(t-1)}, \bm{W})}$ \Comment*[r]{Update $\bm{z}$}
      }
      $z_{j}^{(t)}|\bm{\rho}^{(t)}, \bm{\Theta}^{(t - 1)}, \bm{W} \sim \text{Mult}(1, p_{j1}^{(t)},\hdots,p_{j K_{\max}}^{(t)})$ \;
   }
   \For{$(l, q)$ in $1:K_{\max}$}{
      $(\sigma_{l,q}^{2})^{(t)}|\bm{z}^{(t)}, \bm{W}^{[l,q]} \sim  \text{IG}(\nu^{(p)(t)}_{l,q}/2, SS^{(p)(t)}_{l,q}/2)$  \Comment*[r]{Update $\bm{\Theta}$}
      $\mu_{l,q}^{(t)}|\bm{z}^{(t)}, (\sigma^{2}_{l,q})^{(t)}, \bm{W}^{[l,q]} \sim {N}(\mu^{(p)(t)}_{l,q},\; (\sigma_{l,q}^{2})^{(t)}/n^{(p)(t)}_{l,q})$\;
   }
   \textbf{Store: } $\bm{\Theta}^{(t)}, \bm{z}^{(t)}, K^{(t)} = $ Number of non-zero community labels\;
}
\end{algorithm}

\section{Posterior Inference}
\label{s:inf}

After fitting the models, we now summarize the MCMC samples of model parameters 
after a burn-in period. Letting $B$ be the total number of MCMC iterations
after the burn-in period, the MAP estimator for the community membership allocation vector $\bm{z}$ jointly maximizes the simulated posterior densities across $B$ posterior samples and is defined as
\begin{equation*}
   \widehat{\bm{z}}^{\text{MAP}} \quad = \quad \underset{\substack{\bm{z} \in \{\bm{z}^{(1)}, \hdots, \bm{z}^{(B)}\}}}{\arg\max} f(\bm{W}|\bm{z}, \bm{\Theta})\pi(\bm{z}|\bm{\tau}).
\end{equation*}

For WSIBM, we recommend an alternative approach to estimate $\bm{z}$ \textit{via} a posterior pairwise probability matrix (PPM). The 
off-diagonal entries of the pairwise probability matrix $\bm{M}$ are 
the post-burn probabilities that nodes $j$ and $j'$ are assigned to 
the same community where each $M_{j,j'}\in\bm{M}$ is calculated as $M_{j, j'} = \frac{1}{B}\sum_{b = 1}^B I(z_j^{(b)} = 
z_{j'}^{(b)}|\cdot)$. This technique is preferred over the MAP estimator as it not only utilizes the information from all posterior parameters but also solves the label switching problem by construction (i.e., $M_{j, j'}$ is invariant of label switching). After estimating $\bm{M}$, 
we can then obtain a point estimate of $\bm{z}$ by minimizing the sum of 
squared deviations of its association matrix from the PPM 
\begin{equation*}
    \hat{\bm{z}}^{\text{PPM}} \quad = \quad \underset{\substack{\bm{z} \in \{\bm{z}^{(1)}, \hdots, \bm{z}^{(B)}\}}}{\arg\min} \sum_{j < j'}\left[I(z_j = z_{j'}) - M_{j,j'}\right]^2.
\end{equation*}

This estimation strategy was avoided for estimating $\bm{z}$ for the finite block model (WSBM) as it does not guarantee a fixed value of $K$.

\section{Simulation}
\label{s:sim}

In this section, we validate the Bayesian weighted finite and infinite block models by assessing their community detection performance across various simulation settings. 

\subsection{Simulation Settings}
We simulated small and large undirected networks broadly categorized into two cases having the number of communities set to $K = 3 \text{ and }K=7$, respectively. For the network with $K = 3$ communities, we included $n = 50, 70, 100$ nodes to study the impact of varying sample sizes in community detection. The strengths of within-community interactions are controlled by setting the normal mean weight parameters to $\mu_{l,l} = -3, 0, 3$ for $l = 1, 2, 3$ corresponding to strong negative, weak negative/positive, and strong positive weights, respectively. Moreover, we set the node proportions to $w = n_{l}/n = 0.2, 0.5, 0.3$ allowing different sizes for the three communities, respectively. We further set $\mu_{l,q} = 0$ for all $l \neq q$ assuming weak interactions between communities, as observed in real world data. The variability in weights is controlled by the normal variance parameter $\sigma^2_{l, q} \sim \text{exp}(0.1)$ for all $(l, q) = 1, 2, 3$. Sampling the block variability from an exponential distribution facilitated non-homogeneous variability between different blocks across the entire network. For the larger network with $K = 7$ communities, we included $n = 100, 150, 200$ nodes and set $\mu_{l,l} = -6, -4, -2, 0, 2, 4, 6$ for $l = 1, \hdots, 7$ allowing various degrees of strength for within-community interactions and having node proportions of $w = 0.1, 0.25, 0.15, 0.05, 0.15, 0.1, 0.2$, respectively. The parameters $\mu_{l,q}$ and $\sigma^2_{l, q}$ were set the same as before. For all six simulated cases, we generated $50$ replicates each considering random permutations of $\mu_{l,l}$ $\forall l$ per replicate. 

\subsection{Competing Methods} 
The community detection performance is evaluated across the Bayesian WSIBM and WSBM having different Dirichlet prior settings. We set the concentration parameter to $\alpha = 1$ and $K_{max} = 20$ for the WSIBM model allowing the algorithm to estimate the cluster allocation vector $\bm{z}$ and the posterior density of $K$. On the other hand, we consider three different prior settings for the Bayesian WSBM model by prespecifying the true value of $K$ denoted as $K_{\text{TRUE}}$. In the first setting, the hyperparameters of the Dirichlet prior are set to $\bm{\eta} = (1, 1,\hdots,1)^{\top}$ resulting in a non-informative prior. However, small values of $\eta$ may cause the class sizes $n_l$ to shrink to zero resulting in the algorithm converging to a local optimum \cite{nowicki2001estimation}. Subsequently, setting $\bm{\eta} = (100K_{\text{TRUE}},\hdots, 100K_{\text{TRUE}})$ is the second prior setting recommended by \cite{nowicki2001estimation}. Large values of $\eta$ favor equal class sizes and circumvent the problem of the algorithm getting trapped in an unfavorable region. It may still estimate small class sizes if the data favors so. For the third setting, we set $\eta = 100K_{\text{RAN}}$ where $K_{RAN}$ is randomly (RAN) assigned from $K_{\text{RAN}} \in \{1, \hdots, K_{max}\}$ for the Dirichlet prior hyperparameters. This allows for a more realistic evaluation as the true value of $K$ is generally unknown. The normal-inverse-gamma hyperparameters were set to $\mu_0 = 0, n_0 = 1, \nu_0 = 10, SS_0 = 0.1$ (i.e., non-informative) as default.

The performance of the Bayesian models was compared to frequentist approaches such as the variational EM algorithm \cite{mariadassou2010uncovering}. The R packages \textbf{blockmodels} \cite{leger2016blockmodels} and \textbf{sbm} were used to fit the frequentist WSBM. The frequentist approaches estimate the value of $K$ that maximizes an ICL criterion \citep{biernacki2000assessing}. \textbf{blockmodels} further improved the clustering performance of these models by incorporating absolute eigenvalues spectral clustering \cite{rohe2011spectral} for initializing and reinitializing starting points. To assess community detection accuracy, we evaluate the discrepancy between true community labels $(\bm{z})$ and the estimated community labels $(\hat{\bm{z}})$ \textit{via} two similarity measures: Adjusted Rand Index (ARI) \citep{rand1971objective} and Mutual Information (MI) \cite{steuer2002mutual}. We detailed the definition of these measures in section S1 of the supplementary material. 

\subsection{Simulation Results}
The modeling performance of all the methods was assessed based on their estimation accuracy of $\bm{z}$ and recovering the correct number of communities $K$. Here, we refer to the number of nodes as \textit{node size}. 

\begin{figure}[h!]
    \centering
    \includegraphics[width=1\textwidth]{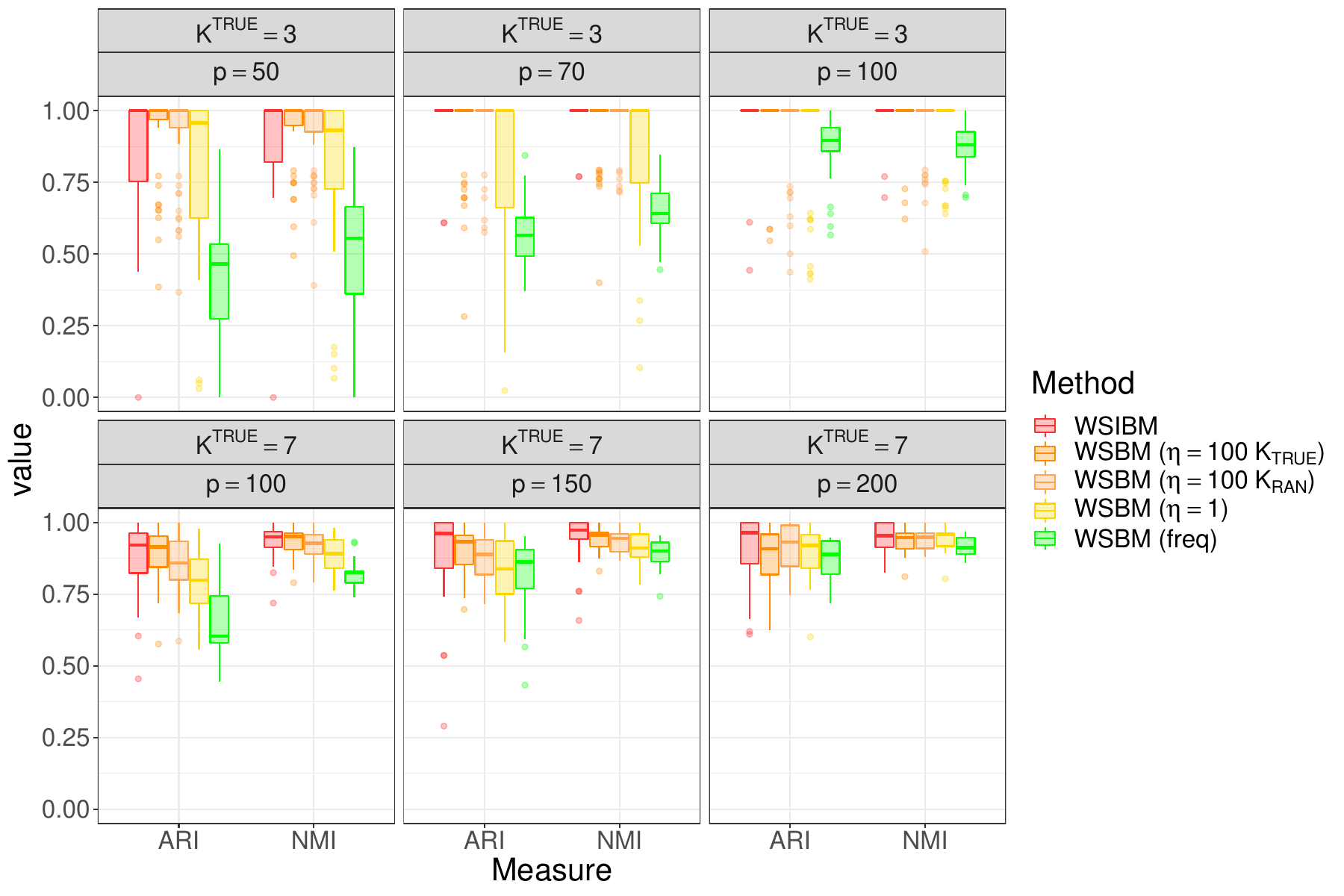}
    \caption{Comparison of clustering accuracy of community membership vector $\bm{z}$ between Bayesian WSIBM (automatic selection of $K$), Bayesian WSBM with prespecified $K$ under various prior settings, and WSBM (freq). ARI and NMI performance metrics were evaluated over $50$ simulated replicates for each of the six cases, where $K_{\text{TRUE}}$ is the true value of $K$ and $n$ is the number of nodes.}
    \label{fig:sim_box}
\end{figure}

In Figure \ref{fig:sim_box}., for the first three cases with $K_{\text{TRUE}} = 3$, Bayesian WSBM with the second prior setting \text{Dir}($\bm{\eta} = (100K_{\text{TRUE}},\hdots, 100K_{\text{TRUE}})$) performed the best on average for a small to moderate number of nodes $(n = 50, 70),$ respectively. This is attributed to the fact that $K_{\text{TRUE}}$ was prespecified as a fixed value of $K$ as well as in the Dirichlet prior. Bayesian WSIBM and WSBM with the third prior setting $\bm{\eta} = (100K_{\text{RAN}},\hdots, 100K_{\text{RAN}})$ performed similarly well for $n = 50, 70$ nodes having close to a $100\%$ clustering accuracy. Bayesian WSBM with the first prior setting (i.e., \text{Dir}($\bm{\eta} = (1,\hdots, 1)$)) performed slightly worse than its other Bayesian counterparts as it shrunk some of the community-specific node sizes to zero in some of the replicates. However, WSBM (freq) severely under-performs having clustering accuracy in the range of $50\%$-$60\%$ for $n = 50, 70$ but improves for large node size ($n = 100$) with $80\%$ accuracy. The poor performance of the frequentist approach is likely due to its inability to account for unequal block variances under small to moderate node sizes. For the cases with $K_{\text{TRUE}} = 7$, WSIBM achieved the best performance even without fixing the true value of $K$. WSIBM estimates $\bm{z}$ \textit{via} PPM instead of the MAP estimator. PPM takes into account all MCMC iterations of $\bm{z}$ post burn-in, while $\bm{z}_{\text{MAP}}$ is just a single iteration of $\bm{z}$ that maximizes the joint posterior density of $(\bm{z}, \bm{\Theta}, \bm{\tau})$. PPM tends to demonstrate the superior estimation of $\bm{z}$ for large node sizes in general (i.e., $n > 100$). WSBM (freq) has comparable clustering accuracy to its Bayesian equivalents when $n = 200$.

\begin{figure}[h!]
    \centering
    \includegraphics[width=1\textwidth]{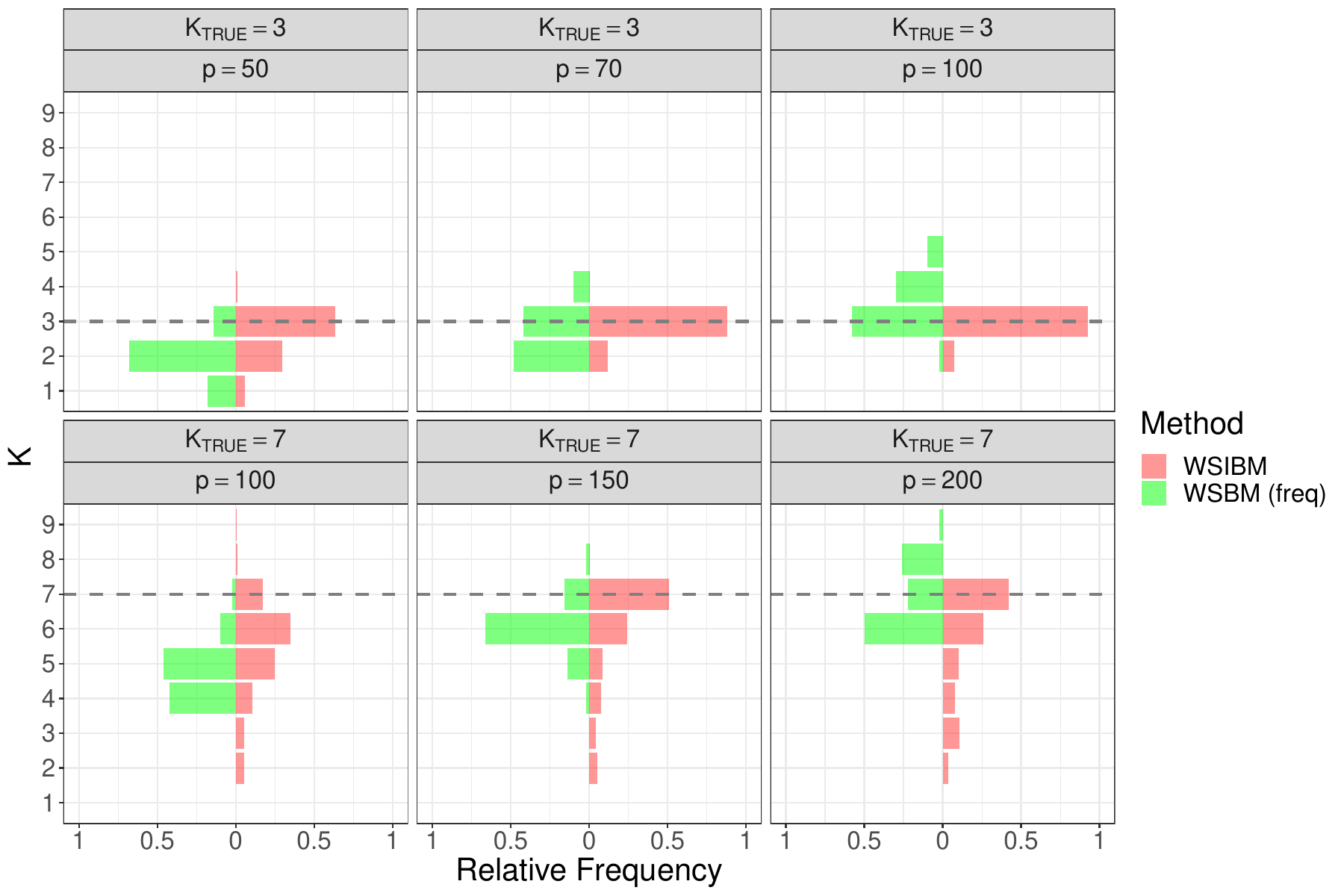}
    \caption{Comparison of the estimation of $K$ between Bayesian WSIBM (automatic selection of $K$) and WSBM (freq). The bars represent the percentage of times a certain value of K was estimated across $50$ simulated replicates per case. The grey dashed line plots $K_{\text{TRUE}}$.}
    \label{fig:sim_bar}
\end{figure}

In Figure \ref{fig:sim_bar}, WSIBM recovers $K_{\text{TRUE}} = 3$ 70\%-95\% of the times out of 50 simulated replicates for small ($n = 50$) to large ($n = 100$) node sizes, respectively. However, it has weaker recovery (20\%-50\%) when $K_{\text{TRUE}} = 7$. Communities with a smaller number of nodes are harder to detect by the clustering algorithm. WSBM (freq) tends to underestimate the value of $K_{\text{TRUE}}$ for small to moderate node sizes and  overestimate $K_{\text{TRUE}}$ for larger node sizes. This corroborates with the weak community detection performance (estimation of $\bm{z}$) \textit{via} WSBM (freq) on the six simulated scenarios. 

In summary, Bayesian WSIBM and WSBM outperform WSBM (freq) when we have weaker signals (e.g., smaller node sizes, higher variability, and unequal block variances). The frequentist method performed similarly as the Bayesian models only for large node sizes. Bayesian WSIBM provides a natural framework to estimate $K$ \textit{via} Bayesian nonparametric stick-breaking processes and solves the label switching problem \textit{via} PPM estimation.

\section{Analysis of rUTI Microbiome Data}
\label{s:res}

\subsection{Background of Data }
We applied WSIBM to a whole genome metagenomic sequencing (WGMS) dataset previously generated by \cite{neugent2022recurrent} for the study of the urinary microbiome in recurrent urinary tract infections (rUTI) in postmenopausal women.  
WSIBM is particularly suitable for this dataset given the high level of zero-inflation ($89\%$) and sparsity exhibited in low biomass microbiomes such as that of the urinary tract. The urinary microbiome data for the $m=75$ postmenopausal women in the study originally included $180$ bacterial species along with their known genera. We filtered out all species that had fewer than seven non-zero counts, which resulted in a total of $n = 99$ species from $41$ genera. We followed the data preprocessing steps described in Section \ref{data_pre} and then conducted analysis \textit{via} WSIBM on the urinary microbiome dataset with $m=75$ postmenopausal women and $n=99$ species-level taxa per subject. 

\subsection{Results}
\begin{sidewaysfigure}
     \centering
     \includegraphics[width = 1\linewidth]{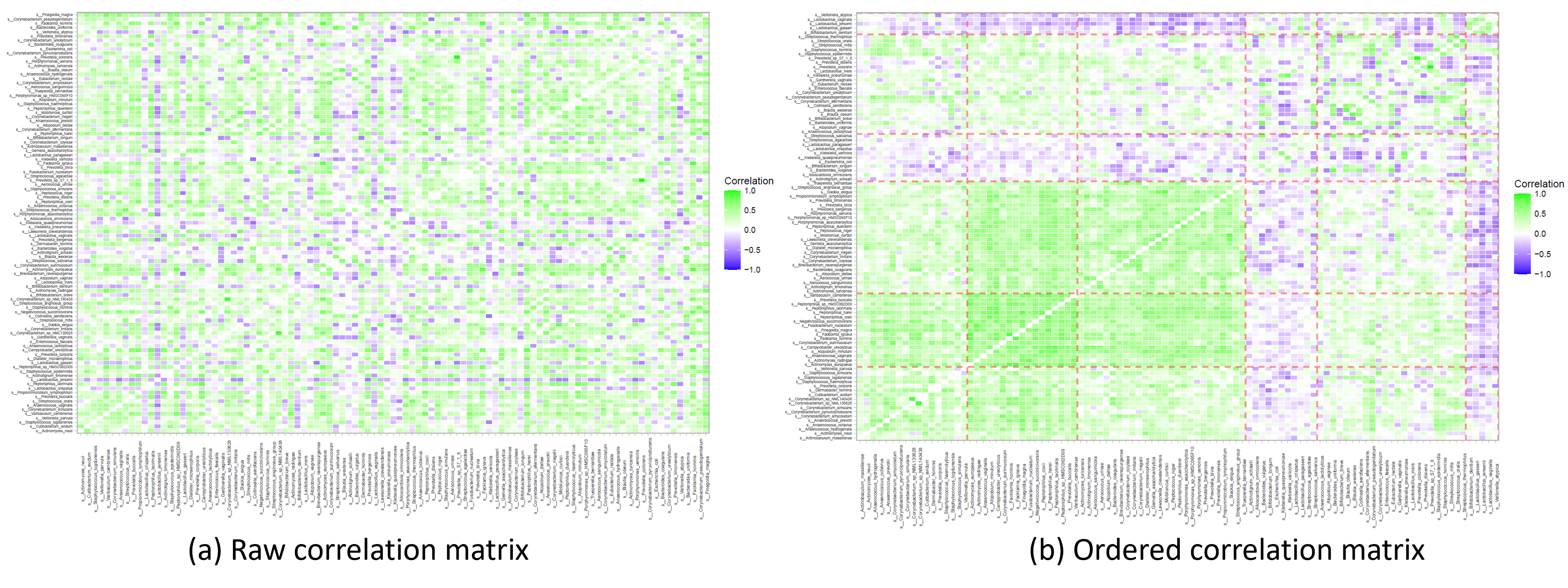}
     \caption{(a) Raw (unclustered) data in the form of taxa-by-taxa correlation matrix; (b) Clustered heatmap of taxa \textit{via} WSIBM. The blocks are separated by red dashed
lines. Strong positive and negative correlations have green and purple tile colors respectively.}
     \label{fig:networks}
\end{sidewaysfigure}

We fit the WSIBM model on the processed data and estimated $K = 6$ communities shown in Figure \ref{fig:networks}. A similar heatmap \textit{via} the frequentist WSBM using the \texttt{blockmodels} package can be found in the Supplementary Materials (Figure S4), which estimated $K = 11$ communities \textit{via} an ICL criterion. Under MCMC algorithm settings, we ran $10,000$ iterations with $B=5,000$ iterations as burn-in. The mean and variance parameter matrices were initialized at $\mu_{l, q} = 0$ and $\sigma^2_{l, q} = 0.1$, for all $(l, q) \in \{1,\hdots, K\}$. The fixed hyperparameters were set to $SS_0 = 0.1$, $\nu_0 = 10$, $\mu_0 = 0$, $n_0 = 1$ while the stick-breaking Dirichlet process parameters were fixed at $\alpha = 0.1$ and $K_{\text{max}} = 20$. 

\begin{figure}
    \centering
    \includegraphics[width = 1\linewidth]{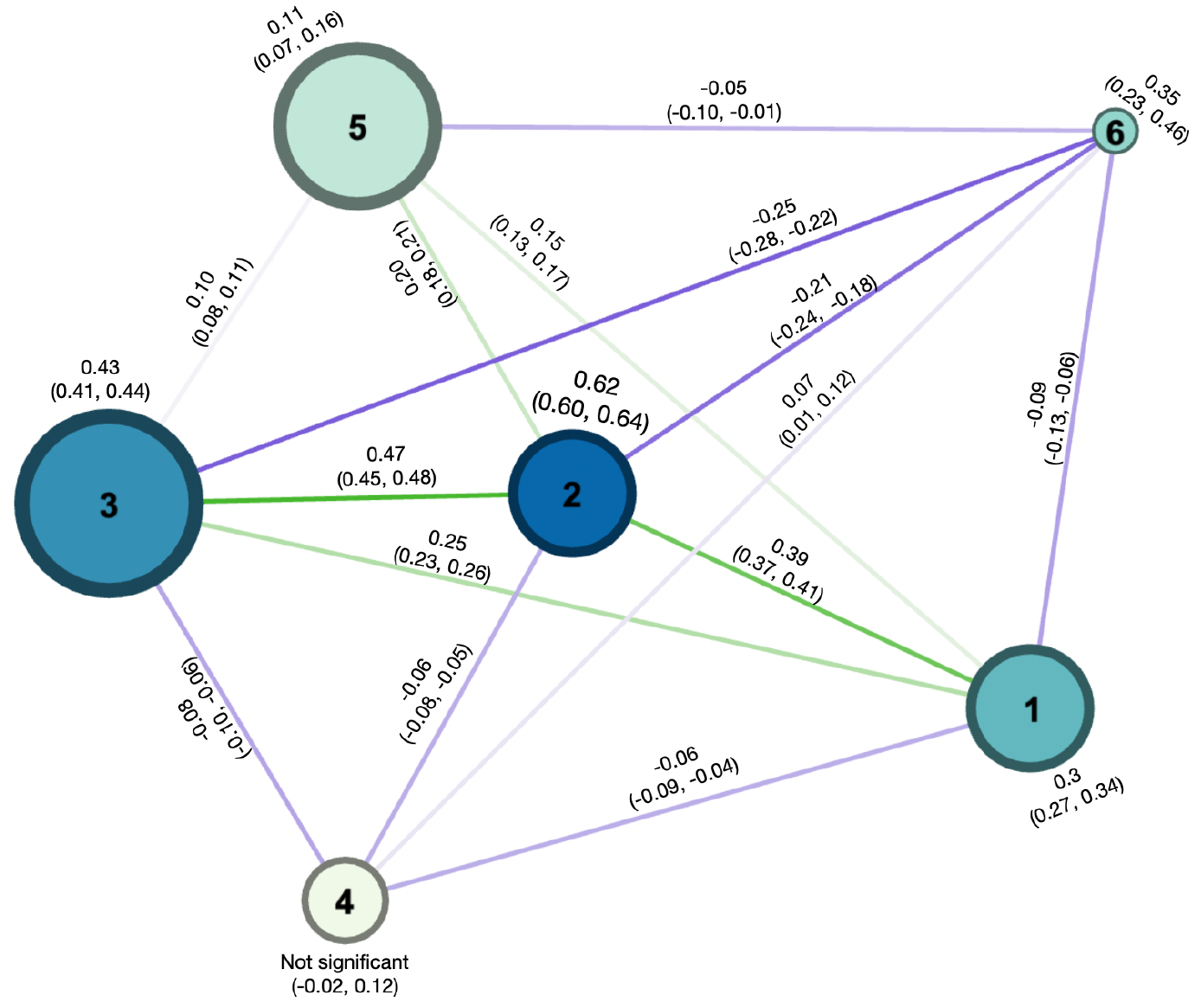}
    \caption{Condensed network plots of the mean correlation weight matrix by community labels and their respective $95\%$ credible intervals. Strong positive and negative correlations have green and purple tile colors respectively. The size of the nodes represents the number of taxa per community.}
    \label{fig:hub}
\end{figure}

In order to ensure a robust estimate of community labeling we ran $100$ parallel chains and obtained $\hat{z}^{\text{PPM}_{100}}$ as an average clustering result across all the chains. We then estimated ARI and NMI clustering measures between $\hat{\bm{z}}^{\text{PPM}_{100}}$ and the $\hat{\bm{z}}^{\text{PPM}}$ estimated over an individual chain i.e. $\hat{\bm{z}}^{\text{PPM}_{1}}$. It was then observed that ARI and NMI estimates were $0.88$ and $0.92$ on average, indicating that  $\hat{\bm{z}}^{\text{PPM}_{100}}$ had converged to a robust estimate of $\bm{z}$ as shown in Figure \ref{fig:networks}. In the network Figure \ref{fig:hub} we provide a network representation of the mean correlation weight matrix by community labels and their respective $95\%$ credible intervals. Here, the nodes represent the communities or within community mean correlations, while, the edges are the between community mean correlations. Community 1 has the largest mean correlation at $0.62$ and is highly positively correlated with communities 1 and 3. This signifies that the taxa in community 2 have strong within-community interactions and also have strong between-community interactions with the taxa in communities 1 and 3. Here, community 3 is the largest community in terms of number of taxa and contributes to roughly $26.3\%$ of the total. On the other hand, community 6 is the smallest community with five taxa only. Community 6 is unique as it has taxa that negatively correlate with taxa from all the other communities while having a strong positive correlation with taxa within the community. Meanwhile, communities 4 and 5 tend to have weaker interactions overall.

\subsection{Biological Interpretation}
\begin{figure}
    \centering
    \includegraphics[width = 1\linewidth]{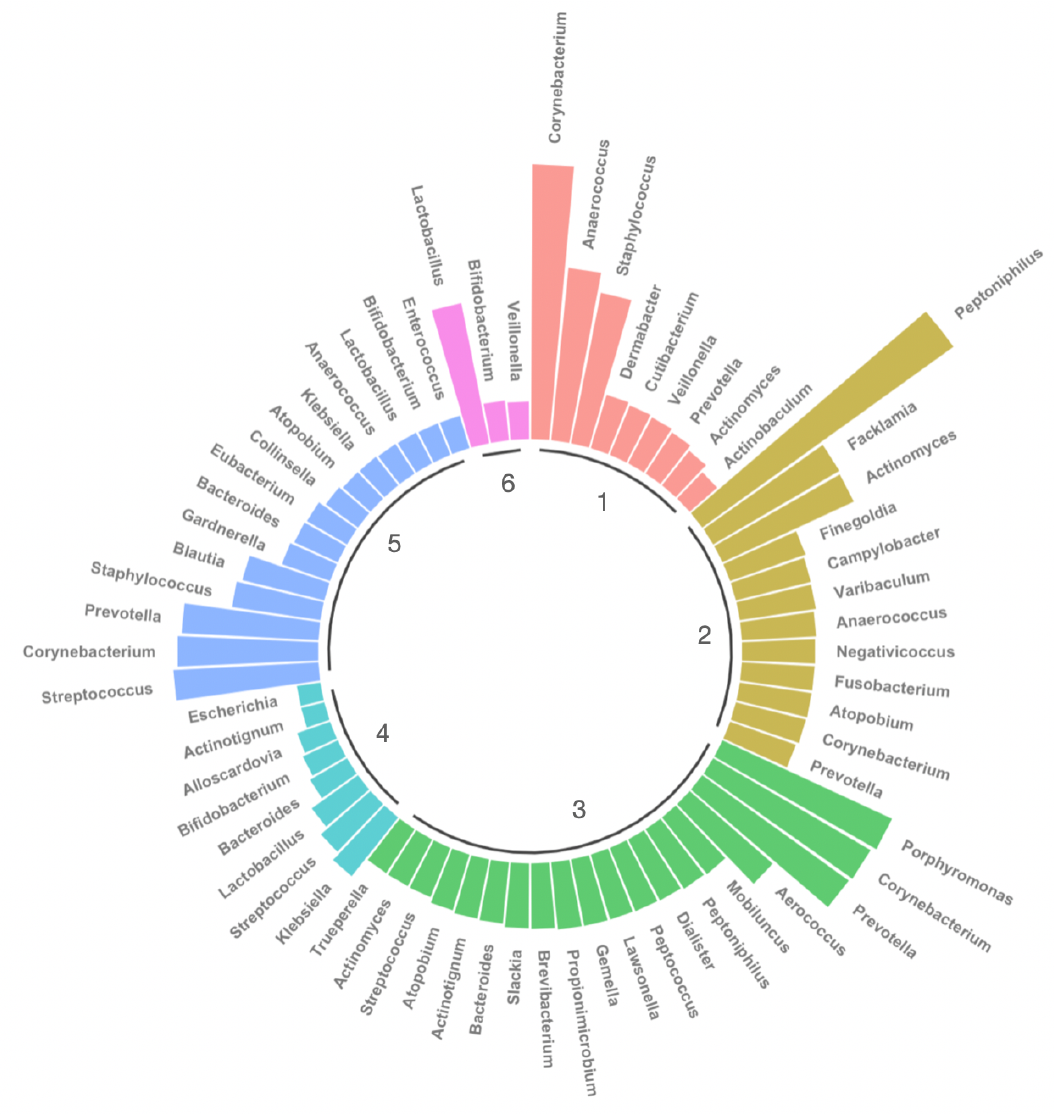}
    \caption{Circular bar-plot of taxa arranged in descending order of absolute weighted nodal strength calculated at a genus level and categorized by the community labels.}
    \label{fig:genus_bar}
\end{figure}

In Figure \ref{fig:genus_bar}, we computed the absolute weighted nodal strength at a Genus level per cluster.
Nodal strength is defined as the sum of all absolute edge weights of a single taxon \cite{hall2019co}, calculated by
\begin{equation*}
    d_{j} = \sum_{j' = 1}^{p} |W_{jj'}|
\end{equation*}
where $d_{j}$ is the nodal strength for taxa $j$. We observed that the top few taxa with the highest nodal strengths tend to influence the community. A table for the maximum absolute correlations per block is provided in the supplement Figure S1. 

In community 1, many of the taxa are known members of the human skin microbiome. \textit{Corynebacterium}, \textit{Staphylpcoccus}, \textit{Dermabacter}, and \textit{Cutibacterium} are all well characterized members of moist skin \citep{chen2018skin}. These skin taxa seem to correlate strongly with dysbiotic taxa similar to those found in communities 2 and 3, both of which harbor known dysbiotic members of the urogenital microbiome such as \textit{Peptoniphilus}, \textit{Anaerococcus}, \textit{Atopobium}, and \textit{Prevotella} in community 2 and \textit{Prevotella}, \textit{Peptoniphilus}, \textit{Atopobium}, and \textit{Streptococcus} in community 3 (Figure \ref{fig:genus_bar}). 

Some genera in community $4$ are known to be mono-dominant taxa. For example, \textit{Lactobacillus} is a known beneficial member of the urinary microbiome and tends to dominate the taxonomic profile when present. \textit{Escherichia}, \textit{Klebsiella}, and \textit{Streptococcus} are known uropathogens and also tend to dominate the taxonomic profile when present. Given that this is the only community with an insignificant within-community correlation, we hypothesized that the clustering is due to the members' tendency to dominate the taxonomic profile. 

Community 5 harbors known members of the urogenital microbiome and seems to be driven by \textit{Streptococcus}, which is considered both a pathogen and a signature of dysbiosis \citep{ceccarani2019diversity, neugent2022recurrent}. This community has the well-characterized \textit{Gardnerella}/\textit{Atopobium} association, which is a signature of vaginal dysbiosis but maybe a distinct ecological signature of dysbiosis form community 2 \citep{bradshaw2006association, hardy2015unravelling, hardy2016fruitful}. This may be evidence that ``dysbiosis" in this microbial niche is not a homogeneous state. There may be multiple dysbiotic taxonomic ecologies.

\section{Conclusion}
\label{s:conclusion}

In this work, we develop a Weighted Stochastic Infinite Block model (WSIBM), which is a nonparametric Bayesian model that jointly estimates community structure as well as the number of communities of taxa using taxonomic abundances from a microbiome dataset. Before fitting the model, our suggested data preprocessing steps such as the MCLR transformation and utilization of a novel correlation method (SPR) designed specifically for microbiome data adjust for non-linearity and zero-inflation in the taxonomic abundance data. Our simulation study demonstrates that our model exhibits superior performance overall compared to the frequentist WSBM model especially when there is a small to moderate number of nodes, higher variability, and unequal block variances. Our implementation of WSIBM on a real urinary microbiome metagenomic dataset generated novel taxa associations and community structure. An interesting finding is that three communities resemble dysbiotic urinary microbiome community states. These data suggest that urinary microbiome dysbiosis may have multiple states, which are not necessarily similar in taxonomic composition.

For future work, we want to scale our model to perform well with big data as the Gibbs sampler becomes computationally unfeasible for large weighted adjacency matrices and high dimensional parameter vectors. We would also like to explore other data applications that have count or nominal edge weights instead of continuous edge weights and provide a general framework that encompasses all the members of the exponential family of distributions.








\clearpage

\bibliographystyle{biom} 
\bibliography{refs}

\label{lastpage}

\end{document}


\maketitle

\section{Evaluation Criteria} 
The simulation study was performed to evaluate the community detection accuracy of $\bm{z}$ and correctly recover the number of communities $K$. We generate 50 replicates of each of the simulated cases and then fit the proposed models. The estimation of $\bm{z}$ is unidentifiable due to the label-switching problem encountered by these models. Hence, we evaluate the discrepancy between true $(\bm{z})$ and the estimated vector $(\hat{\bm{z}})$ via similarity measures that follow a set-matching approach. ARI \citep{rand1971objective, jiang2020bayessmiles} is a corrected-for-chance version of Rand index that does not get affected by the identifiability issues. The ARI can be computed as
\[ \text{ARI}(\bm{z}, \hat{\bm{z}}) \quad=\quad \frac{{n \choose 2}(a + d) - [(a + b)(a + c) + (c + d)(b + d)]}{{n \choose 2}^2 - [(a + b)(a + c) + (c + d)(b + d)]} \]
Here, $a = \sum_{j > j'}I(z_j = z_{j'})I(\hat{z}_j = \hat{z}_{j'})$, $b = \sum_{j > j'}I(z_j = z_{j'})I(\hat{z}_j \neq \hat{z}_{j'})$, $c = \sum_{j > j'}I(z_j \neq z_{j'})I(\hat{z}_j = \hat{z}_{j'})$, and $d = \sum_{j > j'}I(z_j \neq z_{j'})I(\hat{z}_j \neq \hat{z}_{j'})$, respectively, and $I(\cdot)$ is the indicator function. ARI usually takes values between 0 and 1, although it may take negative values. The larger the value of ARI, the better the similarity between $\bm{z}$ and $\hat{z}$. 

We applied another similarity measure that has applications in information theory. Mutual Information (MI) \citep{steuer2002mutual} measures the amount of information one clustering shares with the other. However, the values of MI are unbounded from above and hence harder to compare. Therefore, we implemented a normalized variant of MI \citep{strehl2002cluster} given by
\[\text{NMI}(\bm{z}, \hat{\bm{z}}) \quad=\quad \frac{\sum_{l = 1}^K\sum_{q = 1}^K o_{l,q}\log\left(\frac{o_{l,q}n}{n_l \hat{n}_q}\right)}{\sqrt{n_l\log\left(\frac{n_l}{n}\right)}\sqrt{\hat{n}_q\log\left(\frac{\hat{n}_q}{n}\right)}}\]
where $o_{l,q} = \sum_{j = 1}^n I(z_j = l)I(\hat{z}_j = q)$, $n_l = \sum_{j = 1}^n I(z_j = l)$ and $\hat{n}_q = \sum_{j = 1}^n I(\hat{z}_j = q)$. The values of NMI are bounded between 0 and 1 with higher values attributed to better similarity.

\newpage
\section{Supplementary Figures and Tables}

\begin{figure}[h!]
    \centering
    \includegraphics[width = 1\linewidth]{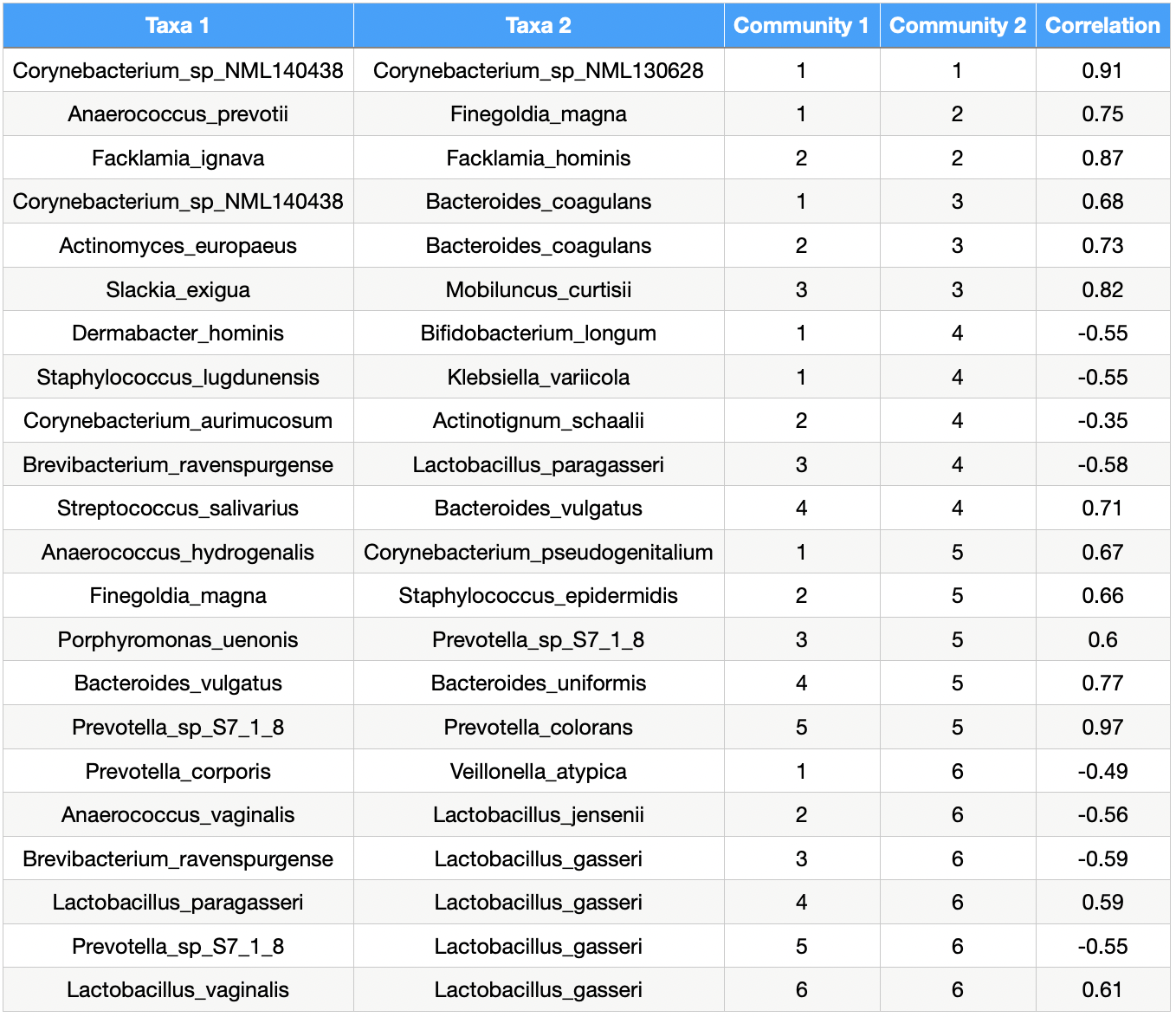}
    \caption{Table of pair of taxa having the maximum absolute correlation per block.}
    \label{fig:hub_table}
\end{figure}

\begin{figure}[h!]
    \centering
    \includegraphics[width = 1\linewidth]{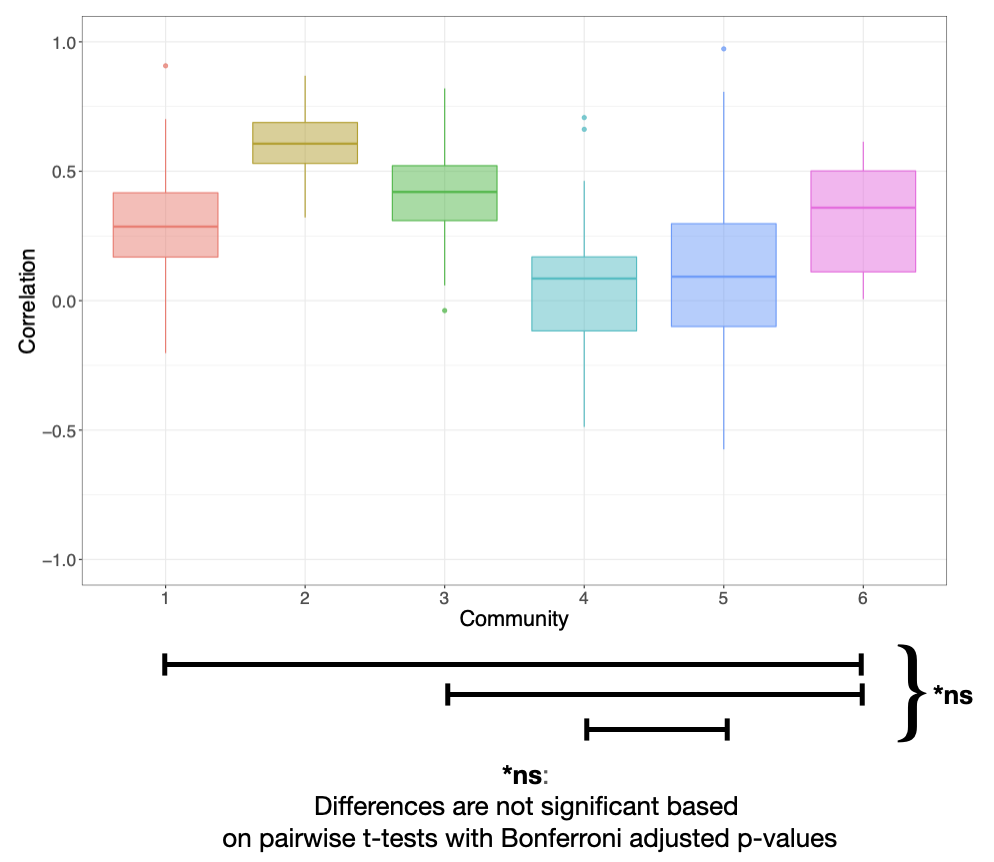}
    \caption{Correlation distribution of each community/diagonal block along with pairwise Bonferroni adjusted t-tests to test for significant mean differences between pairs of diagonal blocks.}
    \label{fig:pairwise_boxplots_diag}
\end{figure}

\begin{figure}[h!]
    \centering
    \includegraphics[width = 1.4\linewidth, angle =90 ]{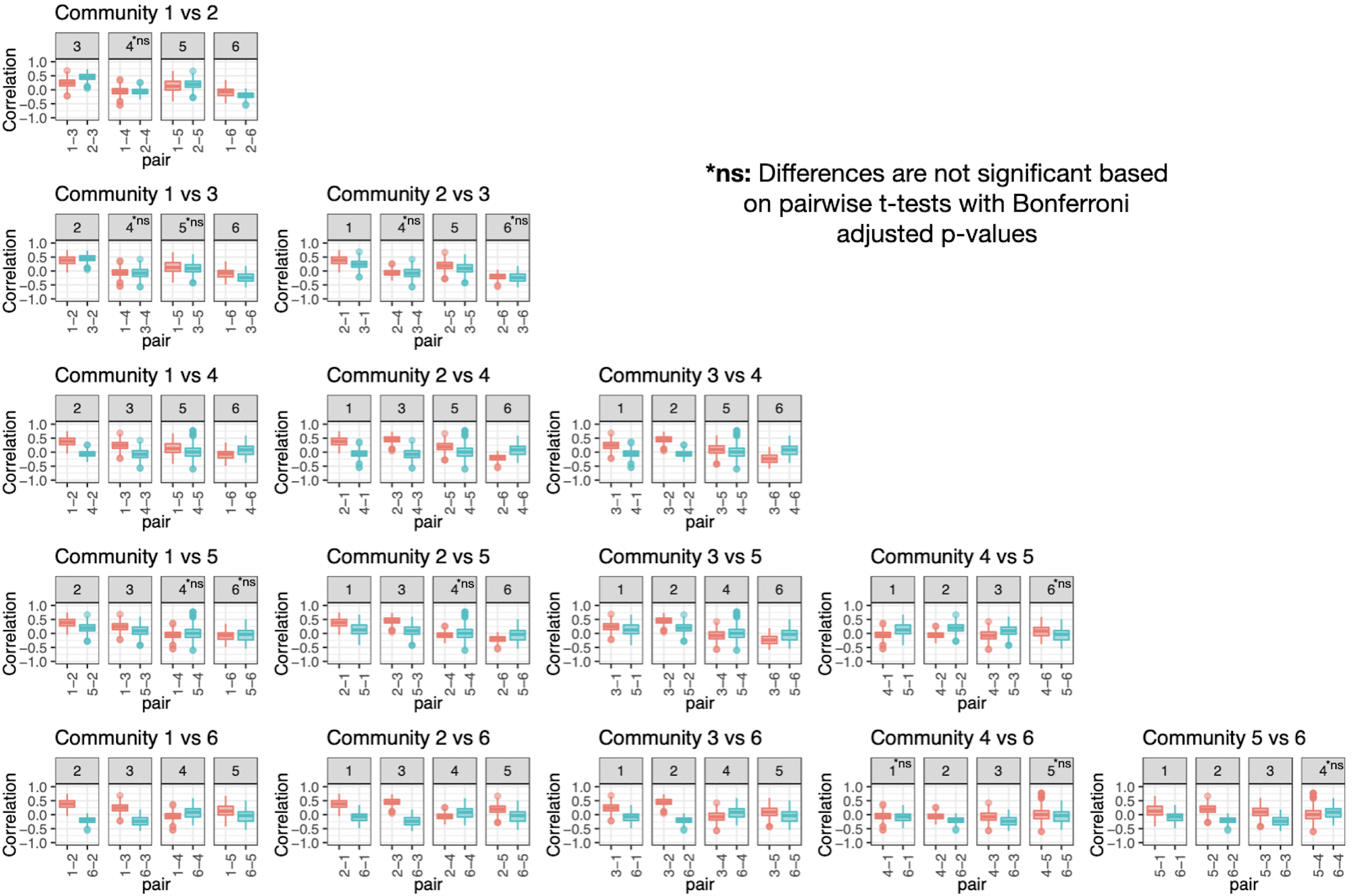}
    \caption{Correlation distribution of each off-diagonal block along with pairwise Bonferroni adjusted t-tests to test for significant mean differences between pairs of off-diagonal blocks.}
    \label{fig:pairwise_boxplots_offdiag}
\end{figure}

\begin{figure}
   \centering
    \includegraphics[width = 1\linewidth]{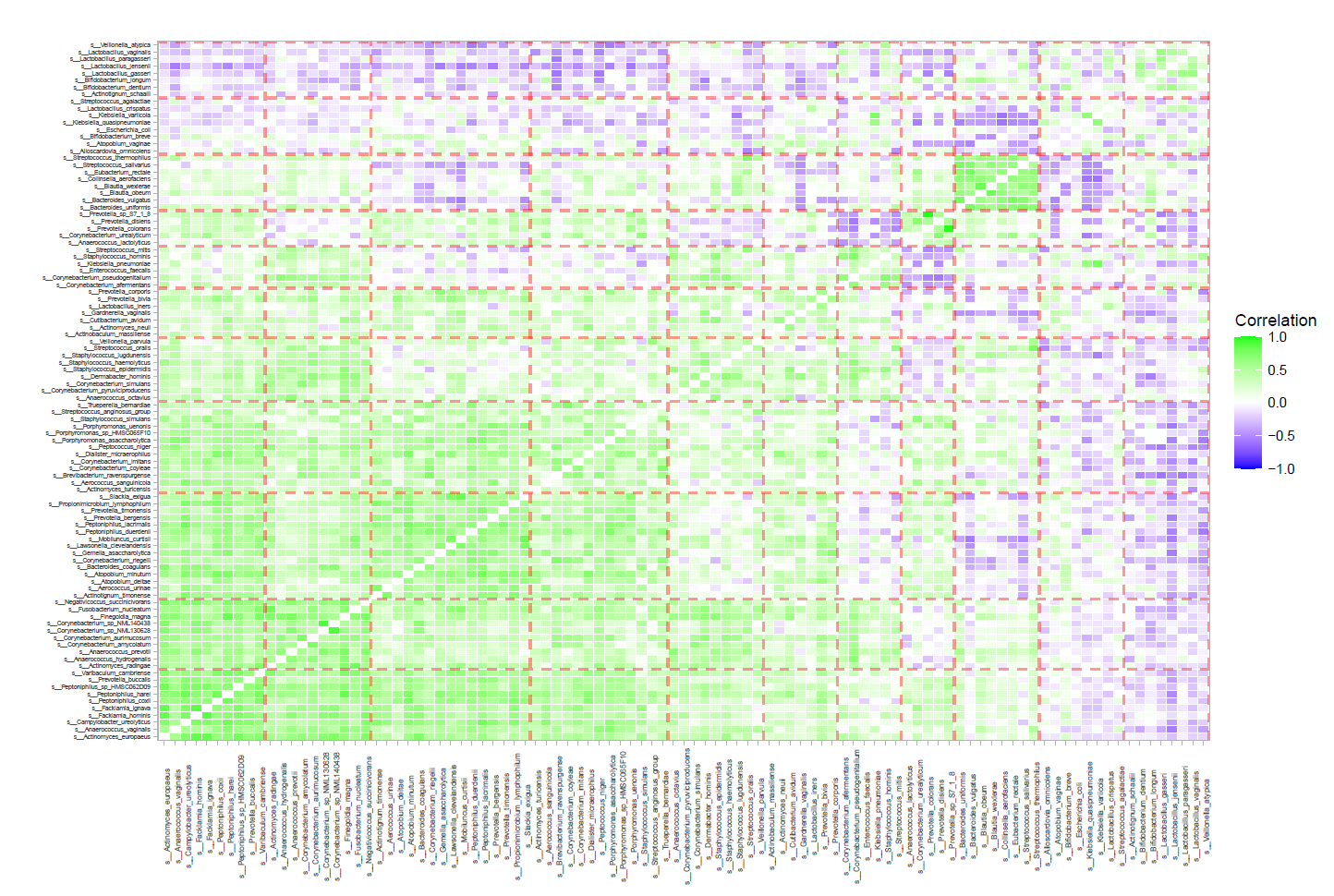}
    \caption{Clustered heatmap of taxa via a frequentist WSBM approach using the \texttt{blockmodels} package. The blocks are separated by red dashed lines. Strong positive and negative correlations have green and purple tile colors respectively.}
    \label{fig:heatmap_freq}
\end{figure}

\begin{figure}
    \centering
    \includegraphics[width = 1\linewidth]{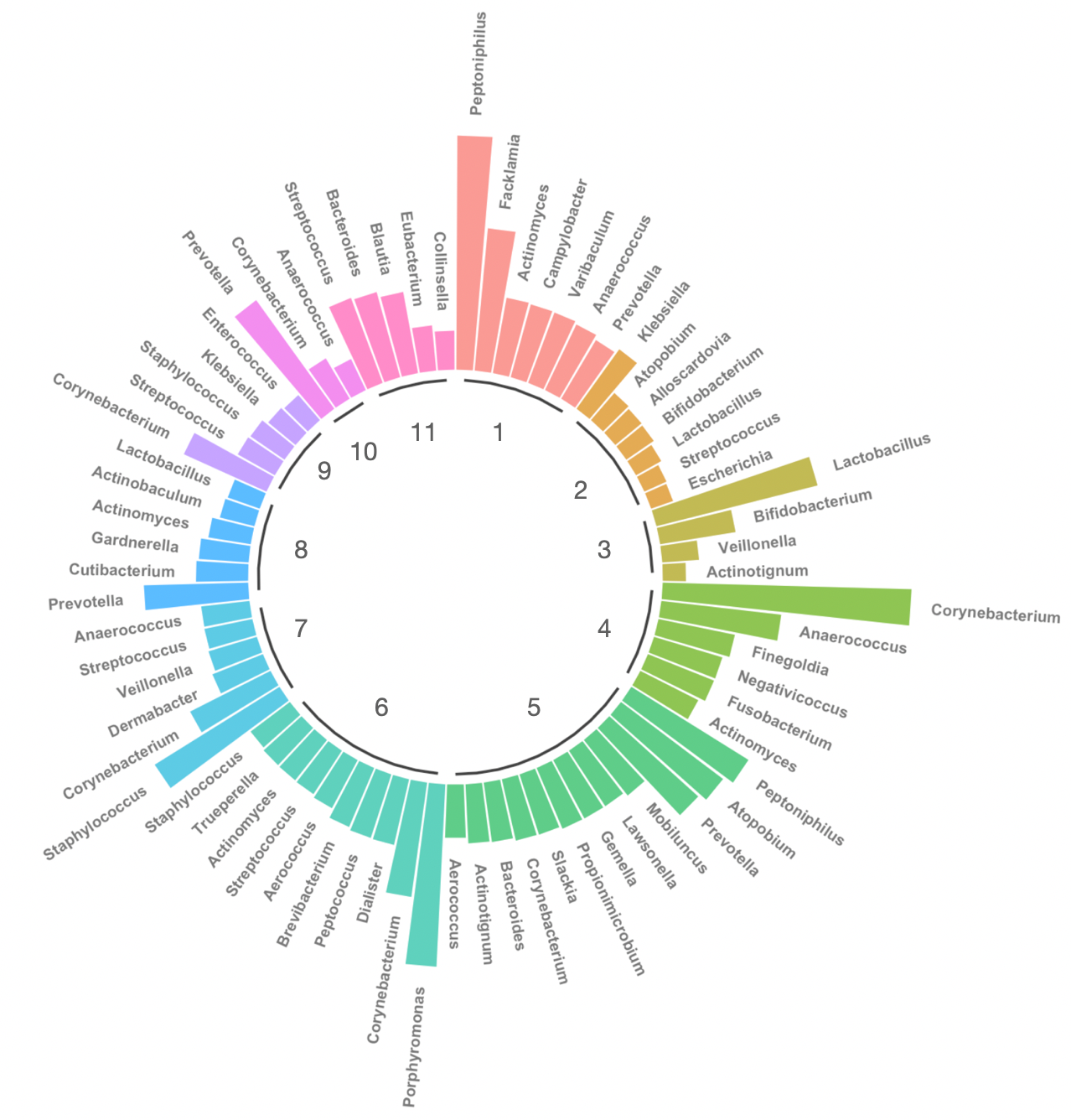}
    \caption{Circular bar-plot of taxa arranged in descending order of absolute weighted nodal strength calculated at a genus level and categorized by the community labels via a frequentist WSBM approach using the \texttt{blockmodels} package.}
    \label{fig:nodal_strength_circular_freq}
\end{figure}

\clearpage

\bibliographystyle{biom} 
\bibliography{refs}